Peer-Reviewed Article
# The challenges of assessing and evaluating the students at distance

Fernando Almeida, PhD
University of Porto & INESC TEC

José Monteiro, PhD
Polytechnic Institute of Gaya (ISPGAYA)

**Abstract:** The COVID-19 pandemic has caused a strong effect on higher education institutions with the closure of classroom teaching activities.  In this unprecedented crisis, of global proportion, educators and families had to deal with unpredictability and learn new ways of teaching.  This article aims to explore the challenges posed to Portuguese higher education institutions and analyze the challenges posed to evaluation models.  To this end, the relevance of formative and summative assessment models in distance education is explored and the perception of teachers and students about the practices adopted in remote assessment is discussed.  On the teachers' side, there is a high concern about adopting fraud-free models and an excessive focus on the summative assessment component that in the distance learning model has less preponderance when compared to the gradual monitoring and assessment processes of the students.  On the students' side, problems arise regarding equipment to follow the teaching sessions and concerns about their privacy, particularly when intrusive IT solutions request the access to their cameras, audio, and desktop.

## The Response of Portuguese Higher Education Institutions to COVID-19

In early March with the declaration of the State of Emergency in Portugal, the higher education institutions (HEIs) had to stop their presential classes. This measure was not applied exclusively in Portugal and was also followed across the European Union countries. As a result of this decision, the Ministry of Science, Technology, and Higher Education (MCTES) stated that Portuguese HEIs must adopt teleworking and distance learning processes (Agência Lusa, 2020).

The biggest challenge of higher education was to adapt to a new distance model that would ensure the continuity of teaching activity. The first question that emerged was how to ensure that all students have the internet and the necessary means (computers or smartphones) to access online classes. Kotowicz (2020) highlights that the problem of lack of technical resources to access online classes is more visible in basic education because about 20% of students do not have this equipment at home, and this number rises to approximately to 1/3 when we consider only public schools. In higher education these numbers are less worrying. Data from Statistics Portugal indicate that only 0.4% of students over 16 do not have the internet (Instituto Nacional de Estatística, 2019). Nevertheless, even though these values are relatively residual, higher education institutions have to consider and support these students.

A second question that emerged was which online platforms should be chosen to teach the classes. It was also questioned whether to follow an asynchronous or synchronous model. The HEIs in Portugal followed different approaches; in most cases, each university dean and polytechnic higher education president only established a set of recommendations, without imposing any model or platform to teach the online classes (Tomé, 2020). The decision was left to the teacher responsible for each course and discipline. Accordingly, classes proceed with the timetable defined by the synchronous classes, but through platforms such as Microsoft Teams or Zoom. Lessons were recorded to later disseminate in the internal information systems of each HEI. Also, the participation on the discussion forums has grown and became an important mechanism for students to expose their doubts. Moreover, email has also gained greater predominance, with a very significant increase of its use to answer questions submitted by students.

A third issue—less discussed but also challenging—was conducting tests remotely. The impact of this issue was very heterogeneous considering the specificities of each course. Two types of questions were raised: (1) how to evaluate students with a concise and strong method in courses whose assessment model was initially defined to be based exclusively on written tests and (2) how to evaluate subjects in which laboratory practice is a component of extreme importance. Although these were two questions asked on opposite sides of the balance, the answer given to them was somehow similar. In the answer to the first question, HEIs sought to conduct written tests through the synchronous platforms or their replacement by essays developed individually or with the cooperation of two or more students; in the answer to the second question, it was mainly recommended the use of simulators to provide to the students the perception of the laboratory activity.

The impact of COVID-19 on Portuguese HEIs cannot be limited to the period of the declaration of the State of Emergency. Even after this period, it will be necessary to implement several safety and hygiene measures, such as the maintenance of social distance, hygiene, and disinfection of spaces and the adaptation of spaces for teaching activities. The Portuguese Ministry of Science, Technology, and Higher Education (MCTES) recommended to maintain the teleworking directives and the adoption of the distance learning. In the same sense, it was recommended to follow distance assessment approaches and avoid traditional evaluation procedures in a classroom to reduce the risk of infection. Furthermore, this distance learning and assessment model is to continue after the COVID-19 pandemic, which offers new perspectives to simultaneously explore classroom teaching with distance learning. There is a political perception that it should be possible to teach with less teaching load, and distance learning can make an important contribution to achieving this goal.

**Mechanisms for Assessing and Evaluating the Students at Distance**

In addition to the way of teaching, the COVID-19 pandemic exposed the challenge of remote learning assessment. Several authors in the higher education field emphasize the importance of formative and summative assessment in the teaching-learning processes (Bognar & Bungić, 2014; Taras, 2008). The formative evaluation aims at the feedback of the whole training process and is carried out throughout the process or action, in all learning situations, on each objective; while the summative evaluation makes a balance of the learning and skills acquired at the end of a training module or unit.  Both evaluation processes are simultaneously responsible for assessing the student's academic performance, measuring his or her evolution throughout classes, and proving the effectiveness of the teaching methodology adopted by the teacher.

The classic models that quantify the student's knowledge based exclusively on written assessment and that dictate the student's approval or disapproval had already been subject to criticism before the current scenario.  However, now with mass remote education, an opportunity arises to rethink these practices and discover new forms of assessment suitable for the virtual environment.  It is also an appropriate time to reconcile these practices to build the future "classroom" teaching.

Tsai (2009) states that remote evaluation should be seen primarily as a form of diagnosis rather than a classification.  In this sense, the evaluation process needs to be continuous and diverse, both in methodologies and tools.  Consequently, the choice of the learning environment is a key element in this process, and tools that offer various forms of interaction should be preferred.

In addition to the adoption of a platform that is complete and innovative, with instruments capable of evaluating different skills and competencies, distance learning is more individualized and student focused.  In a distance learning model, the student has more freedom and space to manage his own time, without the routine being completely focused on his studies.  It will tend to favor productivity, concentration, and motivation.  Furthermore, the adoption of a distance learning model based on an innovative technological platform with diversified learning elements will allow the implementation of the education 4.0 paradigm, in which student learning is personalized and supported by a diverse set of collaborative tools (Almeida & Simoes, 2019; Hussin, 2018).

The distance learning format encourages and facilitates the continuous evaluation process, which changes the traditional paradigms of written tests at the end of each semester.  With this approach, immediate feedback about student performance is provided and the next step in their path is determined.  The activities provided virtually also allow automatic correction, which helps facilitate the teacher's work and reduces bureaucracy.  The teacher's role thus becomes more focused on the student's learning than on their grade.

The spaces for discussions between student and teacher in the forums are incentives for the active construction of knowledge. The student should be encouraged to develop projects that call for involvement with other students and the use of diversified technological tools (e.g., audio, videos, storyboards, etc.). Gamification is another element that can help motivate the student to learn. It increases healthy competition among students and also recognizes progress in student learning. Alomari et al. (2019) state that through gamification it is possible to transform routines of work or study and make people feel more receptive to the tasks and challenges that each situation requires.

Another element that should be promoted in distance education is the interdisciplinarity of content. It is desirable that students have a holistic view of the themes, less focused on the evaluation processes, but instead on how it can be applied in companies and society. In this sense, points of convergence between disciplines should be considered so that the student's knowledge can be applied in multiple areas in an integrated way. The student's evaluation should consider their performance over multiple perspectives and according to the learning outcomes of each class.

### Perceptions from Teachers and Students Community

One of the challenges that was felt at the beginning of the process was that not all teachers were at the same level of preparation for distance learning. Due migration to distance learning was not planned, nor there was time for plan and training on the subject, and some teachers with lower technological affinities felt a little bit lost. The challenge for each HEI IT center was to provide technical support to these teachers. Besides this more formal technical support, the teachers shared informally among themselves various experiences and platform solutions to be adopted in their online classes.

One concern that emerged early on was how the assessment of students should be carried out. There was an excessive focus on the summative component of assessment, as the models adopted and based on a classroom approach were too centralized in this process of competence assessment. Avoiding fraud in the assessment process was a central concern. Several Portuguese HEIs have adopted proprietary and open-source solutions such as ProctorExam, Exam.net, TestWe, and Respondus. These tools allow integration with e-learning Learning Management System (LMS) platforms and offer several mechanisms to verify student identity and environment. Furthermore, they offer other features such as suspicious behavior recording and detection, screen blocking, keyboard shortcuts, screen sharing, 360° smartphone vision. However, despite all these features, none of these solutions are perfect and unequivocally prevent academic fraud.

In isolation, none of the solutions are perfect. Each alternative should be part of an integrative assessment solution that uses both summative and formative approaches. This is also an opportunity to migrate the assessment processes in Portuguese HEIs, excessively focused on the summative component. Furthermore, the solutions proposed by the teachers must receive the agreement of the students. Indeed, students must participate in these processes to ensure that the established assessment model takes into account the development of students' skills and their various rhythms of learning. It is also essential that the technological solution proposed considers the heterogeneity of students' equipment because not all students have the same technical resources at their disposal. Additionally, it is equally important that the assessment is carried out without breaking the students' privacy and comply with the norms defined by the General Data Protection Regulation (EU GDPR) (e.g., implementing a distance learning policy that explicitly indicates the information collected, the reasons for its collection, and for how long and obtaining consent prior to data collection). This is a new area for educational institutions and EU GDPR is gaining greater visibility with distance learning.